\documentclass[english,aps,preprint]{revtex4-1}
\usepackage[T1]{fontenc}
\usepackage[utf8]{luainputenc}
\usepackage{amsmath}
\usepackage{graphicx}
\usepackage{babel}
\begin{document}

\title{Quantum search of a real unstructured database}

\author{Bogusław Broda}

\email{bobroda@uni.lodz.pl}

\homepage{http://merlin.phys.uni.lodz.pl/BBroda}

\affiliation{Department of Theoretical Physics, Faculty of Physics and Applied
Informatics, University of Łódź, 90-236 Łódź, Pomorska 149/153, Poland }
\begin{abstract}
A simple circuit implementation of the oracle for Grover's quantum
search of a real unstructured classical database is proposed. The
oracle contains a kind of quantumly accessible classical memory, which
stores the database.
\end{abstract}
\maketitle
\emph{Introduction}.---Quantum counterparts of some classical algorithms
can substantially speed up various computational tasks \cite{nielsen2010quantum}.
In particular, the famous Grover search algorithm \cite{grover1997quantum},
which is a quantum counterpart of classical search, achieves quadratic
speed-up in terms of oracle queries. Namely, if $N$ is the number
of elements in a search space, to find a single distinguished element
with probability $\mathcal{O}\left(1\right)$, in the quantum case
one should query the oracle only $\mathcal{N}_{\mathrm{Grover}}=\mathcal{O}\left(\sqrt{N}\right)$
times, rather than $\mathcal{N}_{\mathrm{classical}}=\mathcal{O}\left(N\right)$
times, as classically expected.

Originally, the Grover search algorithm was called the database search
algorithm, but the word ``database'' was later dropped. To retain
the term ``database search'' one should distinguish two qualitatively
distinct categories of databases \cite{zalka2000using,williams2001quantum,viamontes2005quantum}:
the real (actual or explicit) database, which is a database in conventional
meaning, and the virtual (abstract or implicit) database, which is
a search space in the meaning of \cite{nielsen2010quantum}. Roughly
speaking, the real database represents data stored in a physical memory
device, whereas the virtual one is not actually a database at all.
Evidently, most of the existing literature is devoted to quantum search
of virtual databases (\cite{reitzner2014two} is one of the few notable
exceptions), instead in the present work we will be exclusively focused
on quantum search of real unstructured classical databases. We should
stress that we leave aside all the potentially important and interesting
issues concerning practical aspects of such quantum searches or their
possible usefulness (for a discussion on such matters, see first of
all \cite{zalka2000using}, and possibly also \cite{williams2001quantum,viamontes2005quantum,reitzner2014two}). 

More precisely, the aim of our work is to present a simple model,
in the form of a quantum circuit, which implements the standard Grover
search algorithm for searching of a real unstructured classical database.
Actually, the only unknown part of the circuit is the oracle, which
is, in the framework of the general Grover search, a black box, but
here it has to be explicitly defined. In particular, the oracle should
contain, as its main component, a kind of quantumly accessible memory,
which stores the database. Architecture of such a memory will be proposed.

\emph{Grover's search for real databases.}---The starting point of
the standard Grover search algorithm is the equal superposition state
\cite{nielsen2010quantum} 
\begin{equation}
\left|\psi\right\rangle =H^{\otimes n}\left|0\right\rangle ^{\otimes n}=\frac{1}{\sqrt{N}}\sum_{x=0}^{N-1}\left|x\right\rangle ,\label{eq:eqsuper}
\end{equation}
where $H$ is the Hadamard operation, and $N=2^{n}$ is the dimension
of the Hilbert space --- the number of elements in a search space.
A particular instance of the search problem can conveniently be represented
by an oracle function $f$, which takes as input an integer $x$,
in the range $0$ to $N-1$. By definition, $f\left(x\right)=1$,
if $x$ is a solution to the search problem, and $f\left(x\right)=0$,
if $x$ is not a solution to the search problem. Next, as a result
of repeated executions of the Grover search operation, the state $\left|\psi\right\rangle $
goes towards the uniform superposition of solution states \cite{nielsen2010quantum}.

In the case of a real unstructured database search, for definiteness,
we will assume, as a starting point, a very natural scheme proposed
in Chapter 6.5 of \cite{nielsen2010quantum}. Namely, suppose we have
a database containing $N$ records, each of length $m$ bits. We will
label these records $d_{0},\ldots,d_{N-1}$. The aim is to determine
where a particular $m$ bit string, $s$, is in the database. First,
let us define the following ``\texttt{LOAD}'' operation \cite{nielsen2010quantum}
\begin{equation}
\left|x\right\rangle \left|0\right\rangle ^{\otimes m}\left|s\right\rangle \frac{\left|0\right\rangle -\left|1\right\rangle }{\sqrt{2}}\overset{\mathtt{LOAD}}{\longrightarrow}\left|x\right\rangle \left|d_{x}\right\rangle \left|s\right\rangle \frac{\left|0\right\rangle -\left|1\right\rangle }{\sqrt{2}},\label{eq:startstate}
\end{equation}
which reveals the data $d_{x}$ located at the address $x$. In the
next step, the second and third registers should be compared, and
if they are the same, then a bit flip is applied to register 4; otherwise
nothing is changed. The effect of this operation is \cite{nielsen2010quantum}
\begin{equation}
\left|x\right\rangle \left|d_{x}\right\rangle \left|s\right\rangle \frac{\left|0\right\rangle -\left|1\right\rangle }{\sqrt{2}}\rightarrow\begin{cases}
-\left|x\right\rangle \left|d_{x}\right\rangle \left|s\right\rangle \frac{\left|0\right\rangle -\left|1\right\rangle }{\sqrt{2}}, & \qquad\textrm{if }d_{x}=s\\
\left|x\right\rangle \left|d_{x}\right\rangle \left|s\right\rangle \frac{\left|0\right\rangle -\left|1\right\rangle }{\sqrt{2}}, & \qquad\textrm{if }d_{x}\neq s.
\end{cases}\label{eq:nielsen37}
\end{equation}

\begin{figure}
\centering{}\includegraphics[scale=0.5]{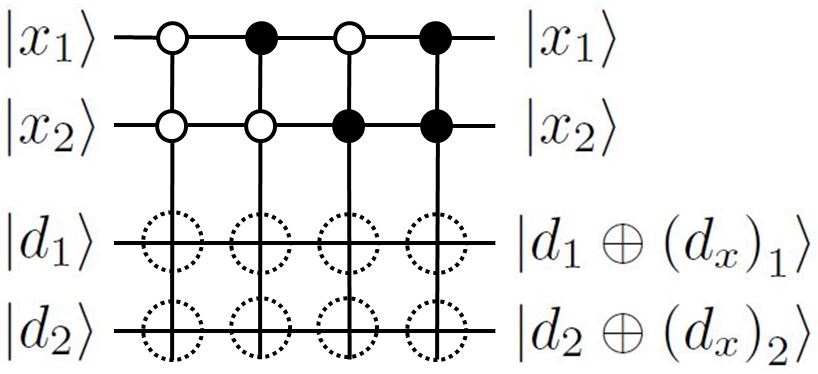}\caption{Architecture of the physical memory device entering the oracle. The
memory consists of an array of $N=2^{n=2}=4$ columns (registers),
which are generalized Toffoli gates. Each column stores a single $m=2$
bit number. In the column $k$ ($k=1,2,3,4$), vertical sequence of
white and black dots is fixed, and it binary represents the number
$k-1$, where the white dot corresponds to 0, whereas the black dot
corresponds to 1, respectively. Vertical sequence of dotted $\oplus$'s
is database dependent, and it binary represents the number (record)
$d_{k-1}$.The dotted $\oplus$ denotes the possibility of the presence
of the actual $\oplus$ in a given place. More precisely, 1 corresponds
to $\oplus$, whereas 0 corresponds to the lack of $\oplus$ (symbolically
``$\cdot$''), respectively. }
\end{figure}

The part of the circuit (actually, of the oracle) performing the operation
(\ref{eq:startstate}) is a kind of memory storing the database. The
memory is supposed to store classical data, in principle, but it should
be possible to access it quantumly. The proposed example architecture
of such a memory is depicted in Fig.~1. The memory consists of an
array of $N$ columns (registers), which are generalized Toffoli gates.
Each column stores a single $m$ bit number. The upper part of the
circuit plays the role of a converter, and it translates the number
$x$ ($x=0,1,\ldots N-1$) from binary to unary numeral system, ``activating''
the column (register) $k=x+1$, whereas the lower part is a proper
memory store. The unitary transformation performed by the memory in
Fig.~1, which depends on the classical database $\left\{ d_{x}\right\} _{x=0}^{N-1}$,
is (compare to \cite{giovannetti2008quantum,reitzner2014two,arunachalam_et_al:LIPIcs:2015:5559})
\begin{equation}
\left|x\right\rangle \left|d\right\rangle \rightarrow\left|x\right\rangle \left|d\oplus d_{x}\right\rangle .\label{eq:dplusdx}
\end{equation}

\begin{figure}
\begin{centering}
\includegraphics[scale=0.5]{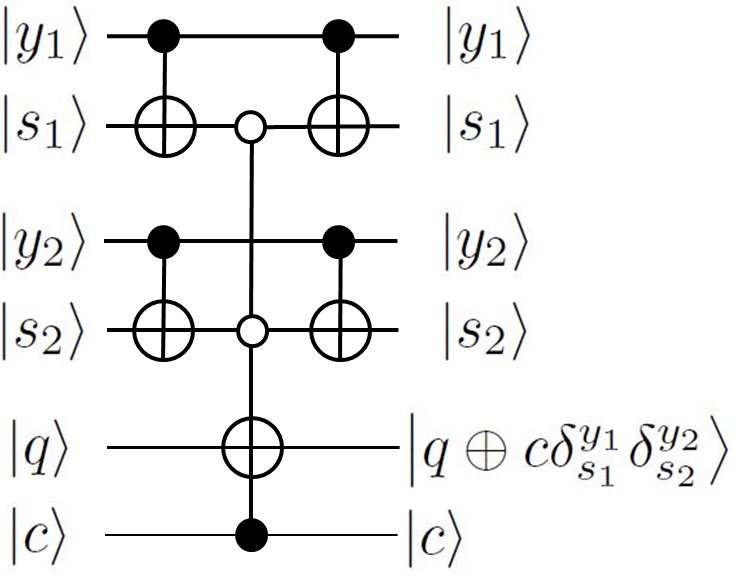}
\par\end{centering}

\caption{For $c=1$, the circuit computes the $s$-dependent oracle function
$f_{\left(s_{1}s_{2}\right)}\left(y_{1}y_{2}\right)=\delta_{s_{1}}^{y_{1}}\delta_{s_{2}}^{y_{2}}$.
Otherwise, i.e.~for $c=0$, the circuit computes the constant function
$f_{\left(s_{1}s_{2}\right)}\left(y_{1}y_{2}\right)=0$.}
\end{figure}

In turn, the part of the oracle which performs the operation (\ref{eq:nielsen37}),
and formally corresponds to the $s$-dependent oracle function
\begin{equation}
f_{\left(s\right)}\left(y\right)=\delta_{s}^{y},\label{eq:stdoraclef}
\end{equation}
is implemented by the circuit presented in Fig.~2. Strictly speaking,
the function (\ref{eq:stdoraclef}) is computed, provided $c=1$,
otherwise, i.e.\ for $c=0$, $f_{\left(s\right)}\left(y\right)=0$.
One should note that for proper functioning of the memory (oracle),
the second register in (\ref{eq:startstate}) should be restored to
its initial value $\left|0\right\rangle ^{\otimes m}$ after each
memory usage (Grover’s iteration). To this end, making use of the
formula $\left|d\oplus d_{x}\oplus d_{x}\right\rangle =\left|d\right\rangle $,
an additional \texttt{LOAD} operation should be performed, which must
not be followed by the computation of the oracle function ($f_{\left(s\right)}\left(0\right)$,
in this case). Therefore, the oracle should consist of the circuits
defined in Fig.~1 and 2, as well as an additional auxiliary CNOT
gate controlling the actual execution of the oracle function $f_{\left(s\right)}\left(\cdot\right)$
(see, Fig.~3). Consequently, the total number of oracle queries is
effectively doubled, but only odd queries contribute to the evolution
of the state $\left|\psi\right\rangle $, whereas even ones restore
the second register of the memory to its proper initial value $\left|0\right\rangle ^{\otimes m}$.
Thus, the whole Grover algorithm should be initialized with the following
set of initial states (Fig.~3):
\begin{equation}
\begin{aligned} & \left|\psi\right\rangle =H^{\otimes n}\left|0\right\rangle ^{\otimes n},\\
 & \left|d\right\rangle =\left|0\right\rangle ^{\otimes n},\\
 & \left|s\right\rangle =\left|s_{m}\cdots s_{1}\right\rangle ,\\
 & \frac{\left|0\right\rangle -\left|1\right\rangle }{\sqrt{2}},\\
 & \left|c\right\rangle =\left|0\right\rangle ,\\
 & \left|1\right\rangle .
\end{aligned}
\label{eq:initial}
\end{equation}

\begin{figure}
\begin{centering}
\includegraphics[scale=0.75]{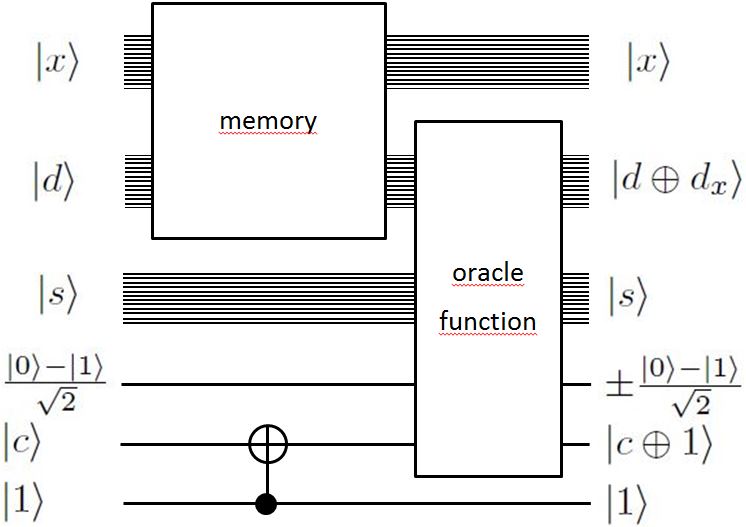}
\par\end{centering}

\caption{Block scheme of the whole oracle. The oracle consists of the physical
memory device proposed in Fig.~1, the oracle function circuit presented
in Fig.~2, and an additional auxiliary CNOT gate controlling the
actual execution of the oracle function $f_{\left(s\right)}\left(\cdot\right)$.}

\end{figure}

\emph{Conclusions.}---In the present work, we have proposed (Fig.~3)
a simple quantum circuit implementation of the oracle for the standard
Grover algorithm adapted for searching of an unstructured real classical
$N$-record ($N=2^{n}$) database. The oracle contains a kind of quantumly
accessible classical memory, which stores the database. It appears
that for proper functioning of the memory, implying the restoring
of the initial value $\left|d\right\rangle =\left|0\right\rangle ^{\otimes n}$,
the number of oracle queries should be doubled, i.e.~$\mathcal{N}=2\mathcal{N}_{\mathrm{Grover}}$.
But one should note that since the oracle is now explicitly defined,
the number $\mathcal{N}$ lose its primary role as a measure of computational
complexity.

As a byproduct of our construction (Fig.~3), we have proposed the
architecture of a physical memory device (Fig.~1), which could be
compared to the qRAM (quantum Random Access Memory) introduced and
extensively discussed in \cite{giovannetti2008quantum,giovannetti2008architectures,hong2012robust,arunachalam_et_al:LIPIcs:2015:5559}.
Actually, only the $\mathtt{LOAD}$ operation, necessary for the realization
of the Grover algorithm, has been defined and quantumly implemented
in our circuit. The lacking $\mathtt{STORE}$ operation, ever performed
only once before the proper Grover algorithm starts, and consisting
in setting the dotted $\oplus$'s (Fig.~1) in appropriate positions,
actual $\oplus$ or ``$\cdot$'', respectively, could possibly be
implemented classically. It is interesting to note that our quantum
memory is not so demanding (exponentially large tensor products) as
anticipated for conventional implementations in \cite{giovannetti2008quantum,giovannetti2008architectures}.
Namely, our memory operates on tensor products of only $\mathcal{O}\left(n\right)$
states, as the bucket-brigade concept of qRAM promises, rather than
on exponent of this number. Incidentally, the circuit of the bucket-brigade
qRAM in \cite{arunachalam_et_al:LIPIcs:2015:5559} operates on exponential
number of states (states are tensored with $\left|1\right\rangle \otimes\left|0\right\rangle ^{\otimes2^{n}-1}$).

\emph{Acknowledgments}.---The author has been supported by the University
of Łódź.

\bibliographystyle{apsrev}
\bibliography{quantum_search_ReDB}

\end{document}